\documentclass[conference,a4paper]{IEEEtran}
\IEEEoverridecommandlockouts
\usepackage{cite}
\usepackage{comment}
\usepackage{amsmath,amssymb,amsfonts}
\usepackage{algpseudocode}
\usepackage{graphicx}
\usepackage{textcomp}
\usepackage{xcolor}
\usepackage{authblk}
\usepackage[linesnumbered,ruled,vlined]{algorithm2e}
\def\BibTeX{{\rm B\kern-.05em{\sc i\kern-.025em b}\kern-.08em
    T\kern-.1667em\lower.7ex\hbox{E}\kern-.125emX}}
\begin{document}

\title{Experimental Evaluation of Air-to-Ground VHF Band Communication for UAV Relays
}

\author{Boris Galkin$^{1}$,
        Lester Ho$^{1}$,
        Ken Lyons$^{2}$,
        Gokhan Celik$^{3}$, 
        and Holger Claussen$^{1,4,5}$
      
}

\affil{$^{1}$ Tyndall National Institute, Ireland; 
       $^{2}$ Irish Defence Forces; 
       $^{3}$ A-techSYN Ltd. \\
       $^{4}$ University College Cork, Ireland;
       $^{5}$ Trinity College Dublin, Ireland\\
\textit{E-mail: boris.galkin@tyndall.ie, lester.ho@tyndall.ie, ken.lyons@defenceforces.ie,} \\ 
\textit{gokhan.celik@a-techsyn.com, holger.claussen@tyndall.ie}}

\maketitle

\begin{abstract}
Unmanned Aerial Vehicles (UAVs) are a disruptive technology that is transforming a range of industries. Because they operate in the sky, UAVs are able to take advantage of strong Line-of-Sight (LoS) channels for radio propagation, allowing them to communicate over much larger distances than equivalent hardware located at ground level. This has attracted the attention of organisations such as the Irish Defence Forces (DF), with whom we are developing a UAV-based radio relay system as part of the MISTRAL project. This relay system will support digital Very High Frequency (VHF) band communication between ground personnel, while they are deployed on missions. In this paper we report on the initial set of experimental measurements which were carried out to verify the feasibility of VHF signal relaying via UAV. In our experiments, a UAV carrying a lightweight Software-Defined Radio (SDR) receiver is positioned at a height of 500 meters above ground, while two 5W transmitters travel in vehicles on the ground. The SDR receiver measures the received signal power, while the Global Positioning System (GPS) coordinates of the vehicles are logged. This is combined to measure the signal pathloss over distance. Our results show that the signal is received successfully at distances of over 50 kilometers away. While the signals still appear to suffer from a degree of obstacle blockage and multipath effects, these communication ranges are a substantial improvement over the ground communication baseline, and validate the use of UAVs to support wide area emergency communication.
\end{abstract}


\section{Introduction}
Unmanned Aerial Vehicles (UAVs) are a disruptive technology that is being adopted in a wide variety of applications, from performing rapid medical deliveries\cite{CNN_2019}, to inspecting hard-to-reach infrastructure such as wind turbines\cite{Shakhatreh_2019}. The telecommunications community have also recognised the potential that UAVs can play in wireless communications applications \cite{Geraci_2022}.  Because they operate in the sky, UAVs experience unobstructed Line-of-Sight (LoS) across very long distances, which means that radio signals coming to and from a UAV node will avoid signal blockage caused by obstacles such as trees, buildings, and even terrain features like mountains. This can allow a relatively small number of UAVs equipped with radio equipment to provide connectivity over a much larger area, compared to an equivalent network of ground-based infrastructure. The capability of UAV-based emergency backup networks was demonstrated by the Alphabet Loon project in Puerto Rico in response to the devastation caused by Hurricane Maria \cite{BI_2017}.

UAV-based radio networks are an attractive solution for organisations that operate in remote areas with unreliable or unavailable infrastructure, such as the Irish Defence Forces (DF). The Irish DF make use of Very High Frequency (VHF) radio communications during their deployments. During overseas missions, DF vehicle convoys may travel hundreds of kilometers on missions that last several days. Due to the distances involved, personnel often find themselves out of range for reliable VHF communications, and have to rely on slower High Frequency (HF) communication. 

The Science Foundation Ireland (SFI)-funded project MISTRAL is developing a network of UAVs that will fly in the air at heights of several hundred meters to several kilometers, and retransmit radio communications between ground personnel, taking advantage of the LoS channel conditions. As part of this project, an experimental measurement campaign is carried out, to determine the propagation conditions of VHF-band radio signals over ground-to-air links. In this paper, we report the results of our first round of measurements. 

Prior experimental work on UAV communication published by the research community tends to investigate UAVs supporting data communication in the higher-frequency bands, such as the ISM or cellular bands \cite{Khawaja_20191}. In \cite{Hovstein_2014} the authors investigated UAV communication via the 2.4 GHz and GPRS bands, and have shown successful video transmission at distances of 50 km.  In \cite{Amorim_2017} the authors investigated UAV connectivity to Long Term Evolution (LTE) networks belonging to multiple operators in a rural environment. In \cite{Al-Hourani_2018} the authors investigated UAV LTE connectivity in a suburban environment, with an exploration of the impact of antenna downtilt on performance. In \cite{Khawaja_2019} the authors investigated short range UAV communication in the $3.1-4.8$ GHz bands. A similar analysis was carried out in \cite{Fuschini_2021} for an urban environment, where building reflections were taken into consideration. The works \cite{Seo_2020} and \cite{Galkin_2021} explored the performance of UAVs connecting to 5G macrocells and small cells, respectively, in urban environments. In \cite{Sanchez_2022} the authors investigated millimeter-wave UAV communication in the 60GHz band, and demonstrated the issues surrounding beam alignment.

To the best of our knowledge, there has been no published set of experimental results on the communication performance of UAVs in the VHF band. Furthermore, with the exception of \cite{Hovstein_2014} and \cite{Amorim_2017} above, published results on experimental UAV communication trials tend to focus on short-range communication scenarios, which are of less relevance to the scenario of basic emergency communication across a wide area. Given this, there is a significant lack of published information on the performance of UAVs acting as long-range VHF relays, despite the prevalent use of VHF communication systems in maritime, emergency response, and military applications. Our work differs from the state of the art by focusing on VHF-band voice communication across distances in the order of tens of kilometers, with UAVs operating at heights of several hundred meters. The results in this paper are therefore of relevance to scenarios where UAVs are used to support emergency communication across long distances.

This paper is structured as follows. In Section II we describe the experimental setup, including a description of the test location and the equipment used. In Section III we present the results of our experiments and discuss our findings. In Section IV we summarise our conclusions and propose future directions for our work.

\section{Experiment Description}
The experimental measurements are intended to mimic the typical behaviour of a convoy of Irish DF vehicles, when performing a humanitarian mission. These convoys travel away from a forward operating base (FOB) towards a destination such as a village. Throughout their journey, they must maintain continuous communication links with the FOB. 

\begin{figure}[htbp]
\vspace{0.5in}
\includegraphics[width=0.45  \textwidth]{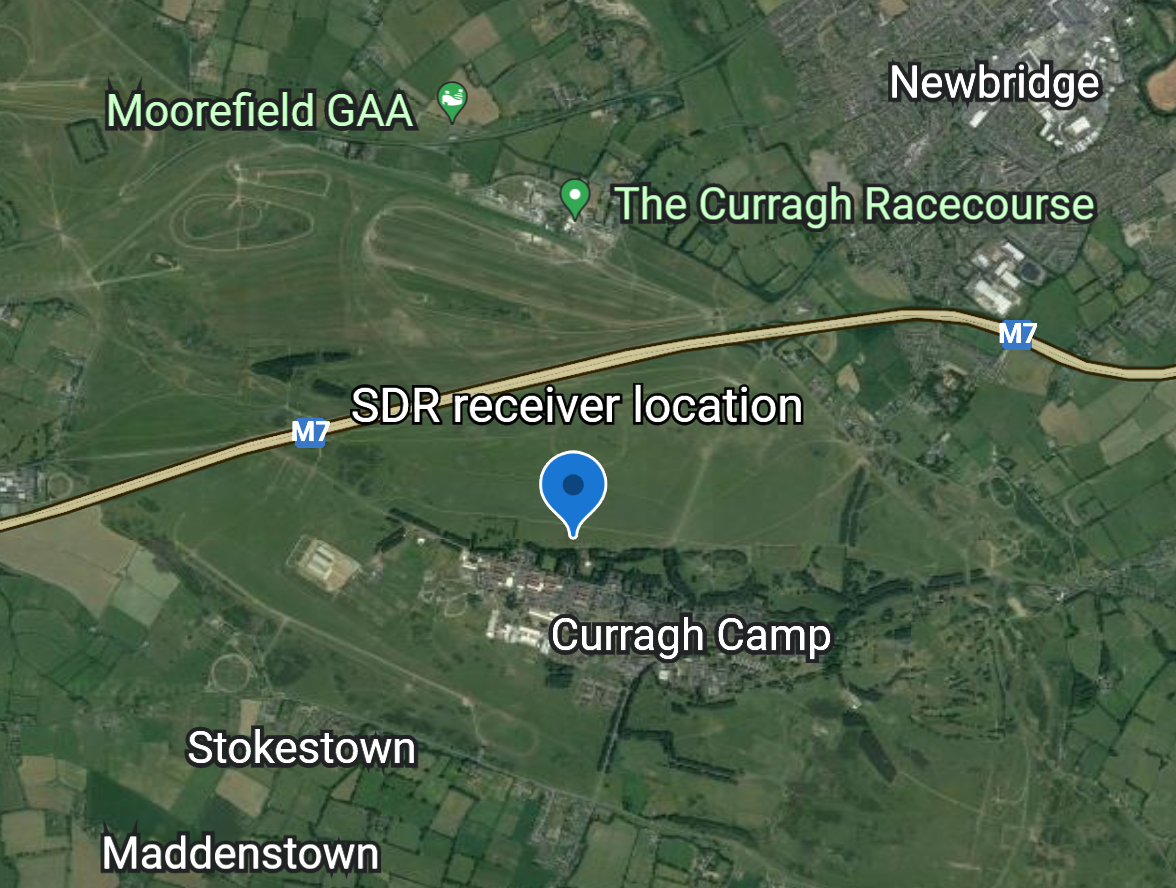}
\caption{Satellite photograph of the Curragh camp and its surroundings, Co. Kildare, Ireland}
\label{curragh}
\end{figure}

To replicate this behaviour, we consider an area around the Curragh military camp located in County Kildare, Ireland, as shown in Figure \ref{curragh}. The camp is situated in a rural region, with fields surrounding it on all sides. The terrain is relatively flat, with gradual height variations of several meters. There are several built-up areas in the region of interest, these are suburban towns and housing estates with low-rise buildings. The M7 motorway passes close to the camp in an approximately east-west heading.

The experiment consists of two measurement phases. In the first phase, we measure the VHF channel at ground level. A VHF transmitter is used to broadcast a signal at regular intervals, from inside a moving vehicle which is travelling around the camp surroundings. The Global Positioning System (GPS) coordinates of the vehicle are logged at the transmitter, while the receiver stores the received signal power with timestamps. In the second phase, the receiver is attached to a UAV which is positioned 500 meters above the camp. Two transmitters broadcast from inside moving vehicles which are travelling in opposite directions on the motorway. As before, the GPS tracks of these vehicles are logged, alongside the timestamped received signal powers.

\begin{figure}[!b]
\includegraphics[width=0.45  \textwidth]{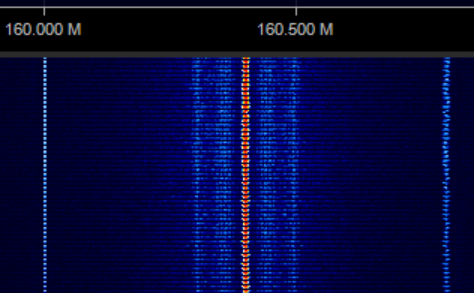}
\caption{waterfall plot of the handset waveform}
\label{waterfall}
\end{figure}

We make use of Motorola DP4801e digital VHF handsets for transmitting the signals under investigation. These handsets have a 5W transmit power with a digital wideband waveform shown in Figure \ref{waterfall}, and are tuned to a center frequency of 160.4 MHz. These handset models are currently in use by the Irish DF which motivates our decision to use them in the experiments; however, they are civilian-grade handsets which are not subject to any export regulations, and are designed for a variety of commercial and industrial applications in the civilian space. As such, the measurement results obtained from using these handsets are widely applicable to other civilian VHF equipment. We choose the 160.4 MHz frequency as it falls inside spectrum that is allocated for maritime and business radio in Ireland, and is therefore commonly used by civilian organisations in outdoor scenarios.

\begin{figure}[htbp]
\begin{center}
\includegraphics[width=0.35\textwidth]{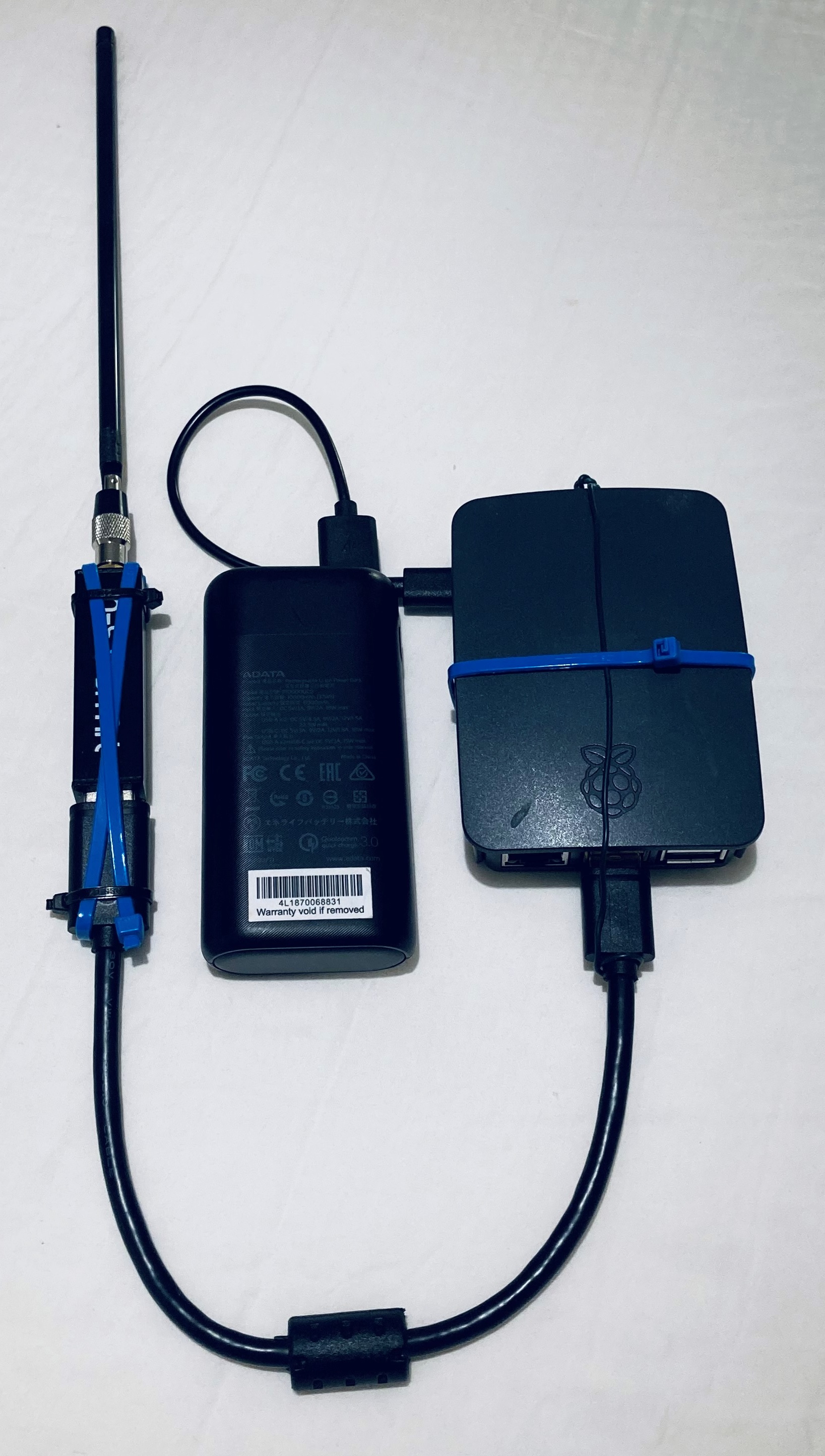}
\end{center}
\caption{Powerbank-powered Raspberry Pi 3 with an RTL-SDR receiver.}
\label{SDR}
\end{figure}

Our Software-Defined Radio (SDR) receiver is an RTL-SDR dongle connected to a Raspberry Pi 3, which is powered by a USB powerbank as shown in Figure \ref{SDR}. The dongle has a telescopic whip antenna which is fully extended to a length of 88 cm. The small form factor and light weight of this setup allows us to attach it to a UAV for the aerial measurements. We use the RTL Python libraries to run a script on the Raspberry Pi which periodically checks the signal power on the 160.4MHz band. From testing, we confirm that the noise floor of the SDR is in the order of $-$105 dBm.

\begin{algorithm}
\SetAlgoLined
 $N = 64$
 
 $n = 1024$
 
 \While{ \textrm{(Recording is not manually terminated)} }{
 
 \For{$i = 1...N$}{
 $s =$ ReadSamples($n$) \algorithmiccomment{Read $n$ samples from SDR}
 
 $PSD_i = (\textrm{FFT}(s)/n)^2$ \algorithmiccomment{Calculate PSD for $i^{th}$ measurement}
 
 }
 
$\overline{PSD}_{\textrm{max}} = \textrm{max}(\frac{1}{N} \sum_{i=1}^{N} PSD_{i}$)
   
Record $\overline{PSD}_{\textrm{max}}$ and timestamp to disk.
}

 \caption{Measurement of Power Spectral Density from SDR}
\end{algorithm}

\begin{figure}[!t]
\includegraphics[width=0.45  \textwidth]{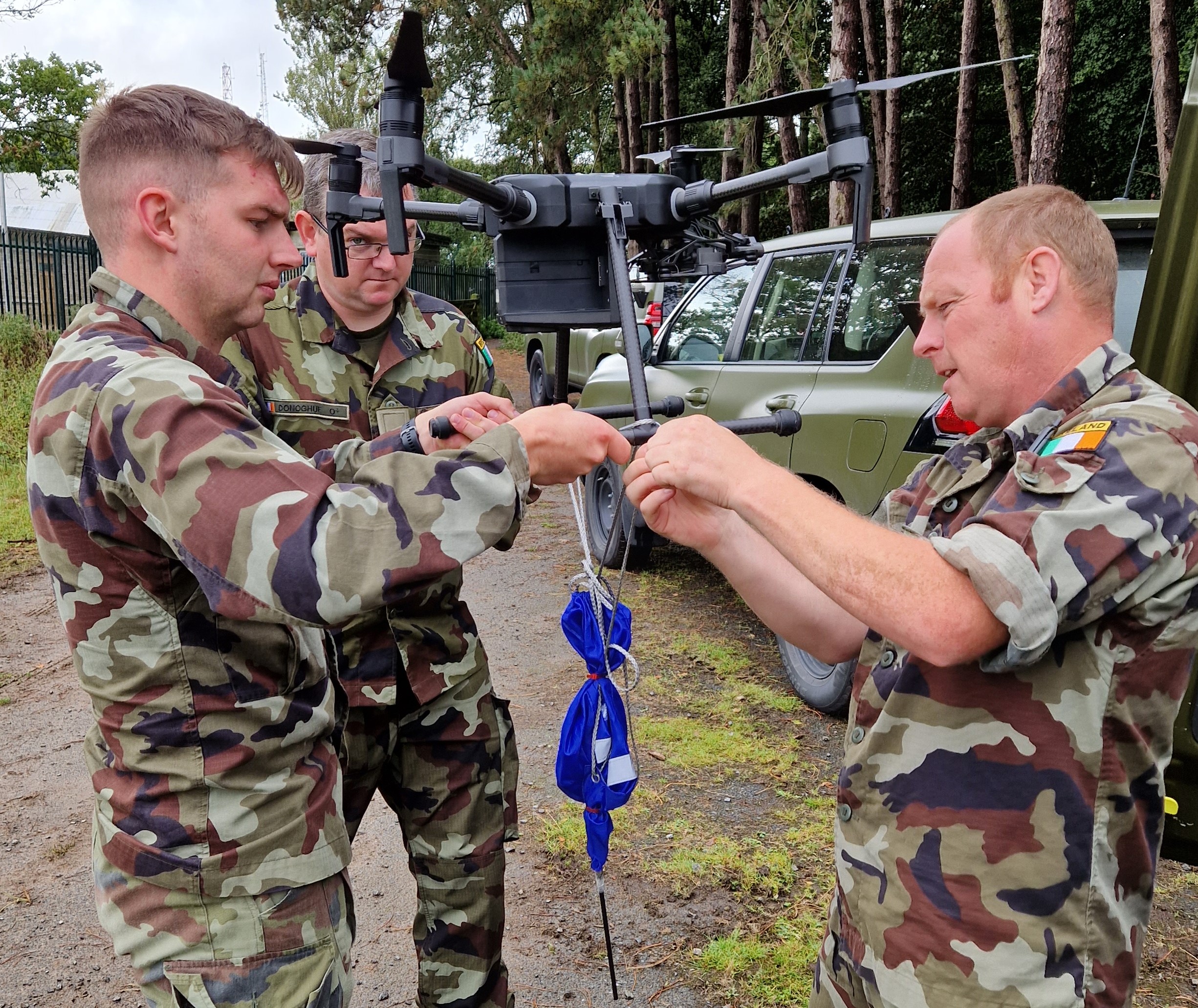}
\caption{DJI Matrice 200 UAV, with the Raspberry Pi SDR suspended underneath.}
\label{drone}
\end{figure}

We use a DJI Matrice 200 UAV to carry the SDR equipment, as shown in Fig. \ref{drone}. The airspace above the Curragh camp is restricted airspace which is exclusively reserved for military use, this allows us to perform our UAV flights above the typical 120 meter legal flight ceiling of civilian UAVs. We position the UAV 500 meters above ground (the hard limit imposed by DJI in their flight software) before the experiments commence, and have the UAV hover at a fixed position during the handset movements.

\section{Experimental Results}

\begin{figure}[!b]
\includegraphics[width=0.45  \textwidth]{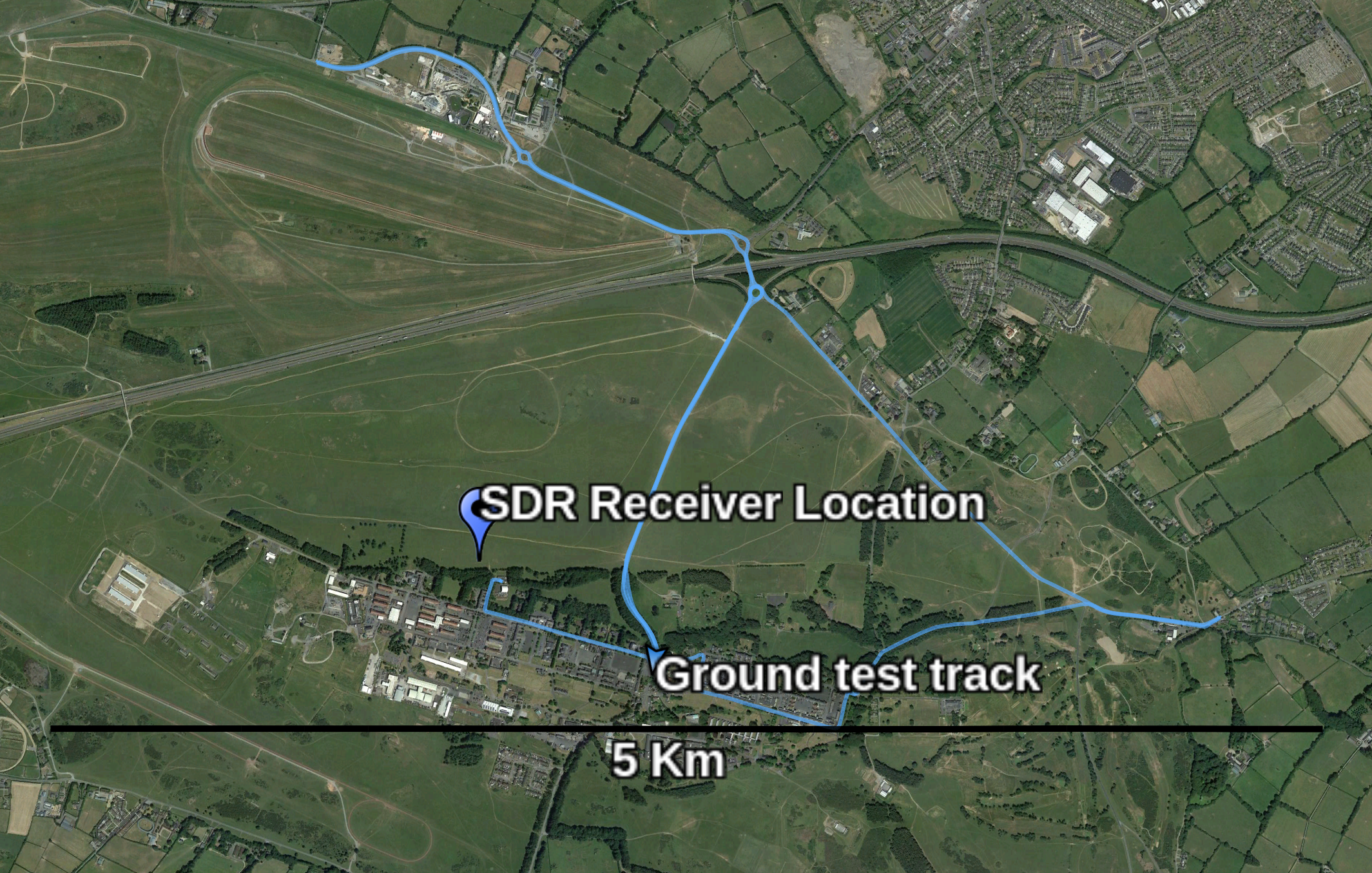}
\caption{GPS trace of the vehicle track for the ground measurements. 5 km line shown for scale.}
\label{groundDrive}
\end{figure}

\begin{figure}[!b]
\includegraphics[width=0.45  \textwidth]{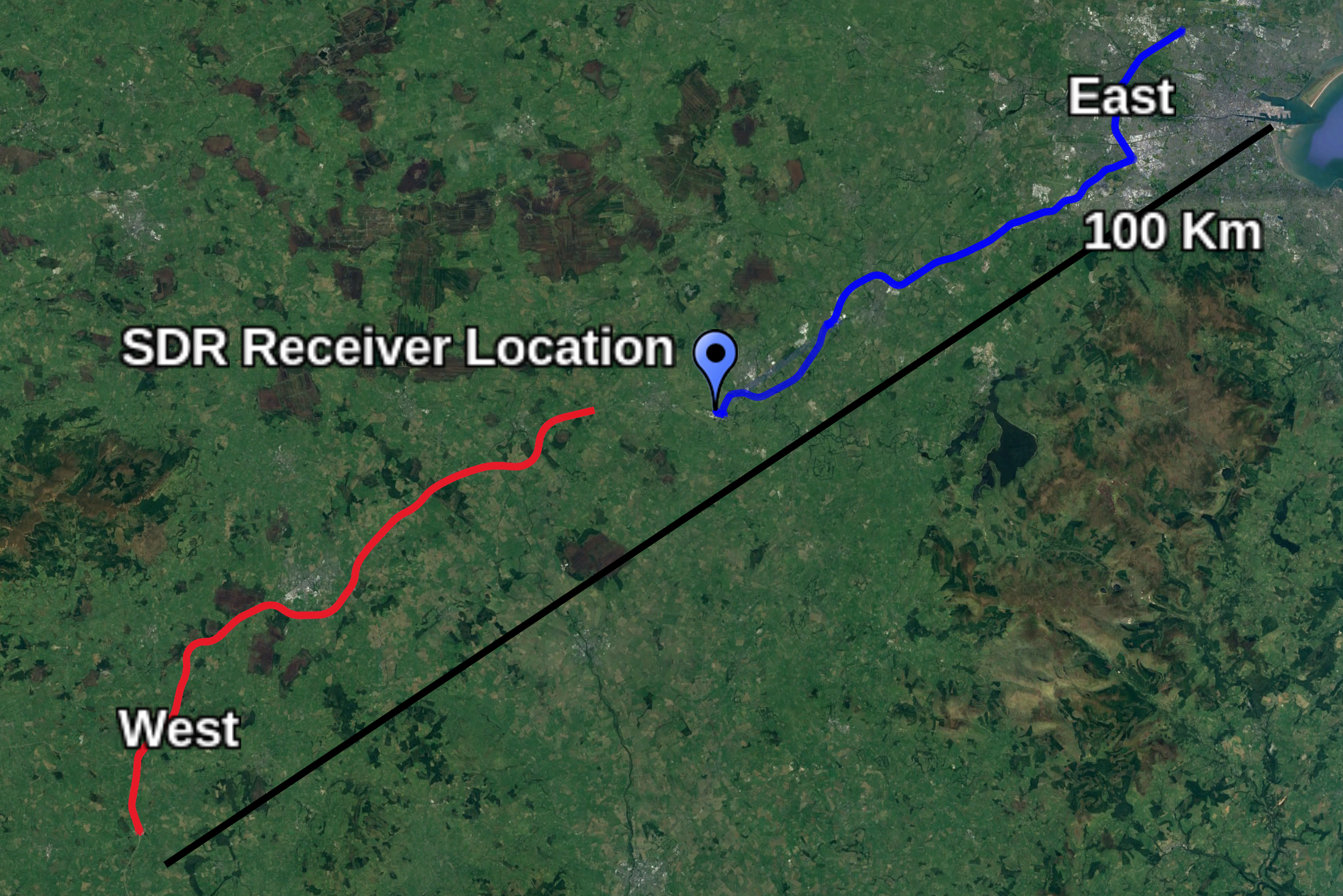}
\caption{GPS trace of the vehicle tracks for the aerial measurements. 100 km line shown for scale.}
\label{airDrive}
\end{figure}

Figures \ref{groundDrive} and \ref{airDrive} show the vehicle paths taken for the ground and aerial measurement parts of the experiment, respectively. For the ground measurements, the vehicle was travelling at speeds of 30$-$50 km/h, whereas for the aerial measurements the vehicles were on the motorway, travelling at speeds between 80$-$130 km/h. The raw data measurement files are available published by us online at \cite{MISTRALDataset}.

We wish to compare our experimentally obtained signal power measurements against the commonly used Free Space Pathloss (FSPL) model \cite{Balanis_2005}, which gives the pathloss in dB as 

\begin{equation}
    PL_{dB} = 10\alpha \text{log}_{10}(d) + 10\alpha \text{log}_{10}(f) + 10\alpha \text{log}_{10}(4 \pi/c),
\end{equation}

such that the received signal power is given in dBm as

\begin{equation}
    Rx_{dBm} = Tx_{dBm} - PL_{dB}.
\end{equation}

$Tx_{dbm}$ denotes the transmit power in dBm, $\alpha$ denotes the pathloss exponent, $d$ denotes the 3D distance between the transmitter and receiver, $f$ denotes the center frequency in Hz and $c$ is the speed of light in $\text{meters/sec}$.  Of these values, $Tx_{dBm}$, $c$ and $f$ are known in advance, and $d$ is found via the GPS logging data. The pathloss exponent is the parameter that is not known in advance, and which we would seek to estimate based on our measurements. The research community generally suggests exponent values between 2 and 4 \cite{Khawaja_20191}, with 2 representing the perfect free space pathloss that can occur under direct LoS conditions, and 4 being for the case of strongly obstructed non-Line of Sight (NLoS) channels. 

\begin{figure}[!b]
\includegraphics[width=0.5  \textwidth]{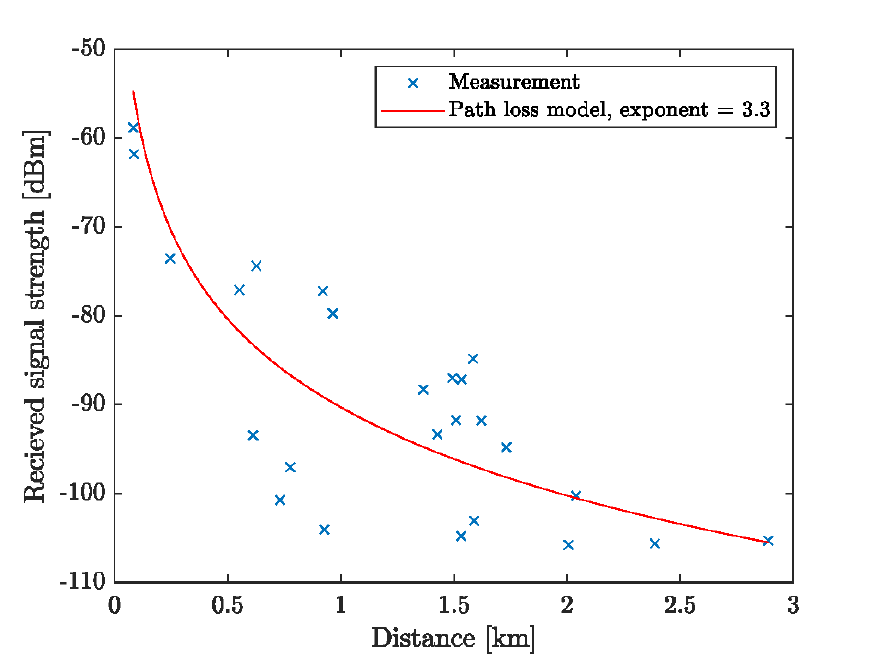}
\caption{Received signal power as a function of horizontal distance, for the ground measurements.}
\label{fig:ground}
\end{figure}

Figure \ref{fig:ground} shows the measured performance for the ground-to-ground channel. We observe that there is significant fluctuation in the received signal strength values, with measurements at 600 meters ranging anywhere from $-$75 to $-$105 dBm. We attribute this to two factors. First, for the environment in question we expect the channels to be largely NLoS, which means that the signal experiences strong multipath effects. VHF band radio signals are also known to experience strong signal diffraction effects, which also contribute to the power fluctuations. Despite the fluctuations, there is a visible signal power decay as the distance increases. Applying the pathloss model given in Eq. (1) we find that this decay can be accurately approximated using the expression, given a pathloss exponent of 3.3. This falls into the range of values typically used by the research community; as expected, the exponent value is closer to the NLoS extreme of 4 identified in prior works such as \cite{Khawaja_20191}, suggesting that at ground altitude wireless communication between two VHF devices tends to behave as an NLoS signal.

\begin{figure}[!b]
\includegraphics[width=0.5  \textwidth]{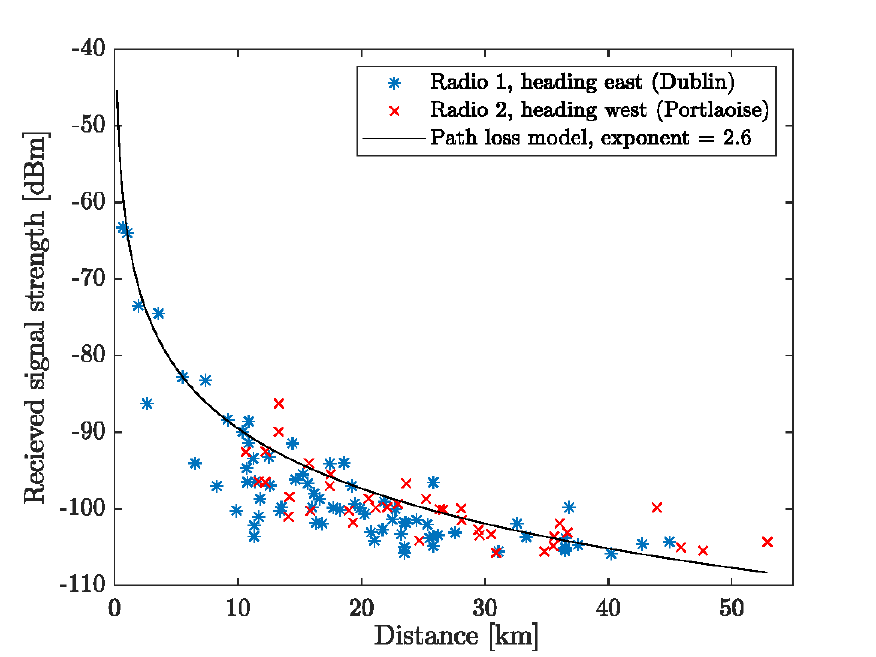}
\caption{Received signal power as a function of horizontal distance, for the aerial measurements at 500m height.}
\label{fig:air}
\end{figure}

Figure \ref{fig:air} shows the measured power as a function of distance for the UAV-mounted receiver. As before, we notice a fluctuation in power due to the pulse-like waveform of the Motorola handset; however, there is a very clear and consistent decay of the maximum received power with increasing distance, for both vehicles. Compared to the ground-to-ground results, the signal is able to travel a much greater distance, with signals being detectable at tens of kilometers away. The maximum detectable distance in our experiment was for the vehicle moving west, at almost 55 kilometers. The east vehicle signal was lost sooner, due to the vehicle entering into a more densely populated environment around the Dublin suburbs, with more signal blockage. Both vehicles experience a signal strength drop that is consistent with Eq. (1), given a pathloss exponent of 2.6. This is a higher value than the free space pathloss of 2, which suggests that there is still some additional attenuation occurring to the signal despite the UAV being positioned 500 meters in the air. Part of this can be attributed to smaller obstructions such as trees and other vehicles on the motorways which cause a degree of signal blockage, albeit not nearly as strong as in the ground-to-ground case. Another contributing factor is the effect of the so-called Fresnel zone. The Fresnel zone is an ellipsoid region surrounding the direct path between a transmitter and a receiver; if obstacles (such as terrain) are found inside this region then we expect them to cause multipath signal reflections and (potentially) destructive interference at the receiver. The width of this region is inversely proportional to the frequency band used; in the cellular bands it may be in the order of a few meters and therefore easily ignored; however, for the VHF band frequencies it may reach a width of over a hundred meters given our horizontal distances. For this reason, we consider that there is some multipath attenuation occurring, even in circumstances where the UAV is expected to maintain LoS to the ground vehicles.

\section{Discussion and Future Work}
In this work we detail the first set of experimental measurements obtained as part of work on creating a UAV relay system to support long-range VHF communications. Our initial measurements have clearly demonstrated a dramatic improvement in the effective communication distance that can be achieved via an aerial relay, due to the LoS propagation that can be achieved at large altitudes. This corroborates the research community's findings regarding UAV communications; however, by experimentally demonstrating signal reception across distances of tens of kilometers using a handheld radio, we validate the idea of using UAV relays to support long-range emergency communication, such as communication used by military and civilian emergency services in a disaster scenario.

Our results raise a number of questions which we intend to explore in subsequent trials. It is clear that the height of the UAV plays a critical part in determining the propagation conditions of the wireless channel and the resulting communication quality; we intend to explore how the channel parameters (namely the pathloss and the random fluctuations) vary at different heights of the UAV. Because we are performing our measurements in reserved military airspace we are capable of operating our UAVs at heights of several kilometers above ground without restrictions; in our future tests we plan to replace our DJI UAVs with ones that are capable of operating at these heights, while carrying several kilograms of radio equipment. This will also allow us to mount a larger variety of radio equipment on-board the UAV, including radio transmitters and signal repeaters. This will give us the opportunity to perform measurement tests where the UAV is acting as a transmitter and the measurement equipment is on the ground. Additionally, we intend to perform tests closer to the Irish coastline, to measure channel conditions between the UAV and sea vessels. This will allow us to explore the capabilities of the UAVs as maritime relay systems.

\section*{Acknowledgements}
This work was funded by Science Foundation Ireland under the MISTRAL project, grant number 21/FIP/DO/9949, as well as the project "GUARD: Drug Interdiction Using Smart Drones" funded under Enterprise Ireland's Disruptive Technologies Innovation Fund, Project ref. DT20200268A. The authors would like to thank the members of the Communication \& Information Services (CIS) Corps of the Irish DF for their assistance with carrying out the experiments.


\begin{thebibliography}{1}

\bibitem{CNN_2019}CNN, “First Drone Delivery of a Donated Kidney Ends with Successful Transplant,” May 2019, Available: https://edition.cnn.com/2019/05/01/health/drone-organ-transplant-bn-trnd

\bibitem{Shakhatreh_2019}H. Shakhatreh et al., "Unmanned Aerial Vehicles (UAVs): A Survey on Civil Applications and Key Research Challenges," in IEEE Access, vol. 7, pp. 48572-48634, 2019

\bibitem{Geraci_2022}G. Geraci, A. Garcia-Rodriguez, M. Mahdi Azari, A. Lozano, M. Mezzavilla et al., "What Will the Future of UAV Cellular Communications Be? A Flight From 5G to 6G," in IEEE Communications Surveys \& Tutorials, vol. 24, no. 3, pp. 1304-1335, third quarter 2022

\bibitem{BI_2017}Business Insider, "Google's parent company has made internet balloons available in Puerto Rico, the first time it's offered Project Loon in the US", Oct. 2017, Available: https://www.businessinsider.com/ap-google-parent-turns-on-internet-balloons-in-puerto-rico-2017-10

\bibitem{Khawaja_20191}W. Khawaja, I. Guvenc, D. W. Matolak, U.-C. Fiebig and N. Schneckenburger, "A Survey of Air-to-Ground Propagation Channel Modeling for Unmanned Aerial Vehicles," in IEEE Communications Surveys \& Tutorials, vol. 21, no. 3, pp. 2361-2391, third quarter 2019 

\bibitem{Hovstein_2014}V. E. Hovstein, A. Sægrov and T. A. Johansen, "Experiences with coastal and maritime UAS BLOS operation with phased-array antenna digital payload data link," 2014 International Conference on Unmanned Aircraft Systems (ICUAS), 2014, pp. 261-266


\bibitem{Amorim_2017}R. Amorim, H. Nguyen, P. Mogensen, I. Z. Kovács, J. Wigard and T. B. Sørensen, "Radio Channel Modeling for UAV Communication Over Cellular Networks," in IEEE Wireless Communications Letters, vol. 6, no. 4, pp. 514-517, Aug. 2017


\bibitem{Al-Hourani_2018}A. Al-Hourani and K. Gomez, "Modeling Cellular-to-UAV Path-Loss for Suburban Environments," in IEEE Wireless Communications Letters, vol. 7, no. 1, pp. 82-85, Feb. 2018


\bibitem{Khawaja_2019}W. Khawaja, O. Ozdemir, F. Erden, I. Guvenc and D. W. Matolak, "UWB Air-to-Ground Propagation Channel Measurements and Modeling Using UAVs," 2019 IEEE Aerospace Conference, 2019, pp. 1-10

\bibitem{Fuschini_2021}F. Fuschini, M. Barbiroli, E. M. Vitucci, V. Semkin, C. Oestges et al., "An UAV-Based Experimental Setup for Propagation Characterization in Urban Environment," in IEEE Transactions on Instrumentation and Measurement, vol. 70, pp. 1-11, 2021

\bibitem{Seo_2020}S. Seo, S. Kim and S. -L. Kim, "A Public Safety Framework for Immersive Aerial Monitoring through 5G Commercial Network," 2020 IEEE Wireless Communications and Networking Conference Workshops (WCNCW), 2020, pp. 1-6

\bibitem{Galkin_2021}B. Galkin, E. Fonseca, G. Lee, C. Duff, M. Kelly et al., "Experimental Evaluation of a UAV User QoS from a Two-Tier 3.6GHz Spectrum Network," 2021 IEEE International Conference on Communications Workshops (ICC Workshops), 2021, pp. 1-6

\bibitem{Sanchez_2022}S. G. Sanchez, S. Mohanti, D. Jaisinghani and K. R. Chowdhury, "Millimeter-Wave Base Stations in the Sky: An Experimental Study of UAV-to-Ground Communications," in IEEE Transactions on Mobile Computing, vol. 21, no. 2, pp. 644-662, 1 Feb. 2022

\bibitem{MISTRALDataset}MISTRAL Project, VHF Drone Radio Measurement Dataset, 2022. Available: https://github.com/MistralProject/VHFDroneMeasurements

\bibitem{Balanis_2005}C. A. Balanis, "Antenna Theory: Analysis and Design", Wiley-Interscience, 2005 

\end{thebibliography}

\begin{thebibliography}{00}
\bibitem{b1} G. Eason, B. Noble, and I. N. Sneddon, ``On certain integrals of Lipschitz-Hankel type involving products of Bessel functions,'' Phil. Trans. Roy. Soc. London, vol. A247, pp. 529--551, April 1955.
\bibitem{b2} J. Clerk Maxwell, A Treatise on Electricity and Magnetism, 3rd ed., vol. 2. Oxford: Clarendon, 1892, pp.68--73.
\bibitem{b3} I. S. Jacobs and C. P. Bean, ``Fine particles, thin films and exchange anisotropy,'' in Magnetism, vol. III, G. T. Rado and H. Suhl, Eds. New York: Academic, 1963, pp. 271--350.
\bibitem{b4} K. Elissa, ``Title of paper if known,'' unpublished.
\bibitem{b5} R. Nicole, ``Title of paper with only first word capitalized,'' J. Name Stand. Abbrev., in press.
\bibitem{b6} Y. Yorozu, M. Hirano, K. Oka, and Y. Tagawa, ``Electron spectroscopy studies on magneto-optical media and plastic substrate interface,'' IEEE Transl. J. Magn. Japan, vol. 2, pp. 740--741, August 1987 [Digests 9th Annual Conf. Magnetics Japan, p. 301, 1982].
\bibitem{b7} M. Young, The Technical Writer's Handbook. Mill Valley, CA: University Science, 1989.
\end{thebibliography}
\end{document}